\documentstyle[epsf,prb,aps]{revtex}
\tightenlines

\newcommand{\ruf}{RuSr$_2$GdCu$_2$O$_8$ }
\newcommand{\ru}{Ru-1212 }
\newcommand{\wn}{cm$^{-1}$ }

\begin{document}

\title{Phonon anomalies and electron-phonon interaction in \ruf  ferromagnetic
superconductor: Evidence from infrared conductivity.} 
\author{A. V. Boris,$^{1,2}$ P. Mandal,$^{1,3}$ 
C. Bernhard,$^4$ N. N. Kovaleva,$^2$
 K. Pucher,$^1$ J. Hemberger$^1$ and A. Loidl$^1$}
\address{$^1$Experimentalphysik V, EKM, Universit\"{a}t Augsburg, D-86135 Augsburg, Germany\\
        $^2$Institute of Solid State Physics, Russian Academy of Sciences, Chernogolovka, Moscow district 142432 \\
          $^3$ Saha Institute of Nuclear Physics, 1/AF, Bidhannagar, Calcutta 700 064, India\\
        $^4$ Max-Planck-Institut f\"{u}r Festk\"{o}rperforschung, Heisenbergstrasse 1, D-70569 Stuttgart, Germany}
\date{\today}
\maketitle
\tightenlines
\begin{abstract}
Critical behavior of the infrared reflectivity of \ruf ceramics is observed
near the superconducting $T_{SC}$ = 45 K and magnetic $T_M$ = 133 K transition temperatures.
The optical conductivity reveals the typical features of the $c$-axis optical conductivity of
strongly underdoped multilayer superconducting cuprates. The transformation of the Cu-O bending
mode at 288 \wn to a broad absorption peak at the temperatures between $T^{\ast }$ = 90 K
and $T_{SC}$ is clearly observed, and is accompanied by the suppression of spectral weight
at low frequencies. The correlated shifts to lower frequencies of the Ru-related
phonon mode at 190 \wn and the mid-IR band at 4800 \wn on decreasing
temperature below $T_M$ are observed.
It provides experimental evidence in favor of strong electron-phonon coupling
of the charge carriers in the Ru-O layers which critically depends on 
the Ru core spin alignment. The underdoped character of the superconductor
is explained by strong hole depletion of the CuO$_2$ planes caused by the
charge carrier self-trapping at the Ru moments. 
\end{abstract}
\section{Introduction}
The layered ruthenate-cuprate compound \ruf  (\ru) is a subject of intense interest as a
unique model system to study the interplay between superconductivity and
ferromagnetism. A ferromagnetic (FM) moment of
the Ru-O layers coexists with superconductivity originating 
from the Cu-O bilayers over a broad temperature range.\cite{Bmuon,Bmeissner,structure1,structure2,Btransport}
Long-range magnetic-order in \ru is established at T$_M$ = 133 K.
Recent neutron-diffraction and magnetization studies\cite{AFMorder,FMorder}
display that in zero external magnetic field the magnetic order of the
Ru moments is predominately antiferromagnetic
along the $c$-axis with spin canting in the $ab$-plane. The net in-plane
magnetic moment is most likely due to the rotations of the RuO$_6$ octahedra.
It is suggested that there is a field-induced spin reorientation,
and that already at moderate fields the order becomes predominately ferromagnetic.
At the same time $dc$ transport and heat-capacity measurements show that \ru
behaves like a typical underdoped cuprate superconductor with the onset of superconductivity
at $T_{SC}$ = 45 K and clearly exhibits a number of features arising from the presence of a
normal state pseudogap. \cite{Btransport}

The importance of $c$-axis charge dynamics for the high $T_C$ superconductivity
still remains controversial.\cite{plasmon1,noplasmon} Many aspects of the $c$-axis transport
properties of superconducting (SC) bilayer compounds like YBa$_2$Cu$_3$O$_{7-\delta}$ (Y-123) have been
explained with a model invoking the onset of inter- and intrabilayer Josephson
junctions.\cite{plasmon1,plasmon2,anomaly3} From this point of view, \ru presents a unique opportunity
to investigate, how the SC Cu-interbilayer coupling propagates through magnetic
Ru-layers.

A more specific question concerns the hole doping of the CuO$_2$
planes required for the superconductivity in \ru compound.
The hole doping obtained from bond-valence summations based on the refined
crystal structure amounts $p\sim$0.4 per Cu atom, and results from an overlap
of the Ru:$t_{2g}$ and the Cu:3d$_{x^2-y^2}$ bands. \cite{structure1}
However, the hole doping of the CuO$_2$ planes derived from 
$dc$ transport and heat-capacity measurements points towards $p$ 
$\sim$0.1.\cite{Btransport} This discrepancy probably 
reflects hole-trapping and/or strong scattering by the ferromagnetic Ru moments.

The present work aims to address these issues by studying
the critical behavior in the infrared (IR) optical conductivity of the \ru ceramics
near the superconducting ($T_{SC}$) and magnetic ($T_M$) transition temperatures. 
The optical conductivity
of \ru reveals the typical features of the $c$-axis optical conductivity of
underdoped multilayer superconducting cuprates: 
Namely, the transformation of the Cu-O bending mode into an additional broad absorption peak
below a characteristic temperature $T^{\ast }$ = 90 K, significantly above $T_{SC}$,
which is accompanied by the relative suppression of the optical conductivity spectral weight
at low frequencies. In addition, a distinct polaron-shaped band at about 4800 \wn
dominates the mid-IR conductivity. On decreasing temperature the increase of the total
spectral weight associated with the intraband transitions is significantly enhanced below $T_M$. 
The correlated shifts of the phonon mode at 190 \wn and the mid-IR band
to lower frequencies below $T_M$ = 133 K provide experimental
evidence in favor of strong electron-phonon coupling of the charge carriers in the Ru-O layers
which is governed by the magnetic order.
\section{Experimental}
Polycrystalline \ru samples were synthesized by solid-state
reaction from high purity RuO$_2$, SrCO$_3$, Gd$_2$O$_3$, and CuO powders,
as described in details elsewhere.\cite{sample,Btransport}
At the final stage of the preparation the sample in the form of a pressed
pellet was annealed at 1060 $^{\circ }$C for 6 days in flowing oxygen and 
was polished to optical quality.  
X-ray diffraction confirms that the sample
is single-phased with a tetragonal $P4/mbm \ (D^5_{4h})$ structure. 
The temperature-dependent magnetization\cite{sample} of the \ru 
samples reveals a magnetic transition temperature  
$T_M$ = 133 K. A large value of diamagnetic shielding is seen below 28 K.
The $dc$ resistivity $\rho (T)$ is similar to that reported recently 
by Tallon {\it et al.},\cite{Btransport} 
and shows the onset of a superconductivity at $T_{SC}$ = 45 K with  
zero resistivity below 32 K. The temperature dependence
of the resistitvity above $T_{SC}$ exhibits two different regimes with
$d\rho /dT \lesssim 0$ for $T_{SC} < T < T^{\ast }$ and $d\rho /dT = const > 0$
for $T > T^{\ast }$ with a crossover temperature $T^{\ast } \simeq$ 90 K.      

Near-normal incidence reflectivities in the far-IR region from 30 to 750 \wn 
were measured using a "Bruker" IFS 133v spectrometer with the newly designed
Ge-coated 6 $\mu m$ mylar beamsplitter. This significantly increased the signal-to-noise
ratio and avoided to merge the spectra at phonon frequencies.
To extend the reflectivity measurements to higher frequencies, a "Bruker"
IFS 66v/S spectrometer was used covering the frequency range from 600 to 16000 \wn, the
higher frequency limit being restricted by calibrating the reflectivity against
the gold-coated sample. The sample was mounted in the "Oxford Optistat"
He bath cryostat for the measurements from 10 to 300 K. 
\section{Experimental results and discussion}   
The reflectivity spectra of \ru for wavenumbers up to 9000 \wn and
for different temperatures from 10 K to 300 K are shown in Fig. 1.
As the temperature decreases from 300 to 10 K, the reflectivity displays a gradual increase,
consistent with an increase of the sample conductivity. Below the SC
transition temperature the far-IR reflectivity is noticeably enhanced.
This increase is reproducible and amounts 3\%.
The reflectivity becomes close to unity at frequencies below 70 \wn in the SC state. 
Comparing to the reflectivity spectra of \ru ceramics reported recently by
Litvinchuk {\it et al.},\cite{ir} our data show a noticeably higher (up to 10 - 20
\%) reflectivity level in the spectral range under investigation. At all temperatures
investigated five optical phonon modes near 128, 151, 190, 288, and 654 \wn are clearly observed.
It is well established\cite{ir,phonons1,phonons2}
that the modes appearing in the spectra of anisotropic
ceramics correspond to phonon modes polarized along $c$-axis.  
The $ab$-plane response from phonons in the reflectivity spectra (left panel, Fig. 1)        
becomes apparent as additional small step-like features at 225, 352, and 520 \wn.
    
To obtain the optical conductivity via Kramers-Kronig analysis, we
used a continuous low-frequency extrapolation of the reflectivity to $R(\omega =0)=1$,
utilizing the formula  $1-const\cdot \omega^{\gamma}$, with $0.5
\leq \gamma \leq 2$.
The lower and upper limits of $\gamma$ correspond to the Hagen-Rubens extrapolation
for the normal state and the extrapolation for the SC state, respectively. 
However, there is no influence of the different extrapolations on
the resulting optical conductivity for wavenumbers $>$ 70 \wn. 
For wavenumbers higher than 16000 \wn, an
extrapolation using a $\omega ^{-4}$ frequency dependence was used.
The real part of the optical conductivity $\sigma _1 (\omega )$ of
\ru for different temperatures is shown in Fig. 2 ($10\leq T\leq 100$: upper panels; 
$100\leq T\leq 300$: lower panels).   
At room temperature the five distinct IR-active phonon modes at 128, 151, 190, 288,
and 654 \wn are superimposed on a featureless electronic background
of 100 $\Omega ^{-1}$\wn, consistent with the room temperature $dc$ conductivity value.
The electronic background extends to higher frequencies and a
broad mid-IR band appears at 4800 \wn.       

The infrared-active phonons observed
in the \ru ceramics (Fig.2, left panels) can be assigned similarly as reported
for the $c$-axis optical phonons in Y-123 crystals.\cite{phonons1,phonons2,phonons3}
Indeed, \ru is structurally similar to the rare-earth cuprate superconductors like
Y-123 except that the Cu-O chain layers are replaced by RuO$_2$ square planar layers.
Ignoring the rotations of the RuO$_6$ octahedra, an approximate structure
is then of tetragonal symmetry $P4/mmm$. 
A comparison with the IR-spectra of Y-123 suggests that
the phonon mode at 654 \wn involves primarily the apical oxygen vibrations,
whereas that at 288 \wn is related to the Cu-O bending
mode which involves vibrations of the oxygen ions of the CuO$_2$ planes.
We suggest the three low-frequency modes at 128, 151, and 190 \wn to
be assigned to the displacements of Cu, Gd, and Ru, respectively. 
Indeed, the Cu-related mode at 128 \wn is close in frequency to
the in-plane Cu vibrations at 104 \wn in Y-123 crystals.    
The eigenfrequency of the Gd mode at 151 \wn 
agrees with that of rear-earth (R) related phonon mode in R-123 cuprates, 
where it is strongly R-dependent decreasing from 193 \wn in Y-123 to 162 \wn in Er-123 
in accordance with mass ratio of the R ion. The assignment is confirmed by shell-model
calculations for \ru with a $P4/mbm$ space group, reported recently
by Litvinchuk {\it et al.}.\cite{ir} These model calculations also suggest
that the mode at 190 \wn involves primarily Ru-ion
vibrations against the Cu ions.    

In order to determine the resonant frequencies of the observed phonon modes
and the mid-IR band parameters at all temperatures,
we applied a quantitative dispersion analysis 
by fitting both the reflectivity and optical conductivity spectra. 
The model dielectric function consists of a sum of Lorentzian functions,
accounting for the contributions from the low-frequency phonon modes at 128, 151, and 190
\wn and the mid-IR band as well. Additionally, the almost featureless electronic background 
at frequencies below 800 \wn was included in a Kramers-Kronig consistent way
into the fitting procedure using a sum of five broad Lorentzian bands with halfwidths more than 150 \wn.    
We have found that the two high-frequency phonon modes, mostly related to oxygen
vibrations, are noticeably asymmetric in shape. 
The asymmetric phonon feature at 654 \wn can be 
described by the Fano profile (see e.g. Ref.\cite{Fano}), given by 
$ \sigma(\omega)=i\sigma_{0}(q-i)^2 \cdot (i+x)^{-1}$, with
$x=(\omega^{2}-\omega_T^{2}) / \gamma\omega$, $\gamma$ 
and $\omega_T$ are the line-width and the resonant frequency, and
$\sigma_{0}$ denotes the amplitude which is related to the
oscillator strength as  $ S= 4\pi \sigma_{0}(q^2-1)\gamma \omega_T^{-2} $.
The dimensionless Fano parameter $q=-1/\tan(\Theta/2)$ is 
a measure of the asymmetry of the peak and for $\Theta=0$ a Lorentzian line shape is recovered.
The asymmetry of the band at 654 cm$^{-1}$ 
yields $\Theta \sim $ 0.19 at all temperatures in accordance
with the Fano-profile description. 
A single resonance description fails for the strongly asymmetric phonon feature 
around 288 \wn, and we used two Lorentzian profiles to describe this
band. The complex nature of the asymmetric structure involving the 
Cu-O bending mode at 288 \wn will be verified by a detailed
analysis of its temperature dependence.  
   
Analyzing the temperature dependence of the optical conductivity of
\ru, we emphasize its striking resemblance of the $c$-axis 
conductivity of strongly underdoped SC Y-123 cuprates.   
Figure 3 shows the most apparent feature related to the onset of superconductivity
which is the transformation of the bending mode at 288 \wn (light solid curve in Fig. 3(a):
$T$ = 100 K) to a broad band centered at 308 \wn (heavy solid curve in Fig. 3(a): $T$ = 10 K). 
Figure 3(b) shows the detailed temperature dependence of the peak position of the asymmetric
structure involving the 288 \wn  mode and of the Lorentzian damping factor of the fitting curve, 
which properly describes the low-frequency wing of the structure.
The transformation of the band profile starts at around 
90 K, close to $T^{\ast }$ defined above, being well below the magnetic transition
temperature $T_M$ = 133 K, but also well above $T_{SC}$ = 45 K. 
The most pronounced changes are observed at $T_{SC}$, 
as can be clearly seen from the temperature dependence of the band position
(open squares in Fig. 3(b)). This behavior involving an anomaly of the Cu-O
bending mode, accompanied by the formation of an additional
broad absorption band at phonon frequencies, is well studied in the $c$-axis 
IR-conductivity of underdoped layered cuprates.\cite{plasmon2,anomaly3,anomaly1,anomaly2}
The mechanism of this phenomenon is a subject of ongoing
discussions.\cite{plasmon1,noplasmon,plasmon2,anomaly2}
The most favored explanation is based on a model where
the bilayer cuprate compounds are treated as a superlattice of 
inter- and intrabilayer Josephson junctions.\cite{plasmon1,plasmon2,anomaly3}
In this model the additional absorption peak is related to the
transverse bilayer plasmon, while the anomalous softening and the loss of the spectral weight of the
Cu-O bending mode are explained to be due to the drastic changes of the local electrical fields
acting on the in-plane oxygen ions as the Josephson current sets in.
The gradual onset of the anomalies above $T_{SC}$ is explained due to the persistence of 
a coherent superconducting state within the individual copper-oxygen bilayers, 
whereas the steep and sudden changes at $T_{SC}$ occur when the macroscopically coherent superconducting
state forms. This effect is strongly dependent on the doping level and is not observed
in the optimally doped and overdoped Y-123 samples.  
For the strongly underdoped YBa$_2$Cu$_3$O$_{6.45}$, the transverse optical
plasmon and the bending mode merge, forming a single highly-asymmetric peak. 
Comparing the temperature dependence of the asymmetric peak
in \ru at 288 \wn with the data reported by Munzar {\it et al.} for  YBa$_2$Cu$_3$O$_{6.45}$
(see Fig. 3 of Ref.\cite{plasmon2}) we find a striking similarity.
The present observation suggests a strongly underdoped character 
of the weakly coupled cuprate biplanes and a strong anisotropy of the \ru compound.
At the same time, the pronounced anomaly of the bending mode observed right at $T_{SC}$
implies that the \ru is a bulk superconductor at lower temperatures.
This experimental evidence is important seeing the onset of a bulk superconductivity
in \ru is still very much disputed.\cite{Bmeissner,Zhu}

Another well-known feature of the underdoped cuprate superconductors is
a suppression of spectral weight (SW) at low frequencies already above $T_{SC}$
due to the opening a normal state pseudogap.\cite{pseudogap}
This feature is not clearly observed in \ru. On the contrary, as shown in the left panels of Fig. 2
the electronic background at low frequencies increases continuously as the temperature
is lowered from 300 K to 100 K, and remains almost constant below 100 K.
The weak suppression in the electronic background conductivity below 400 \wn for $T <$ 100 K
is shown in Fig. 4. However, the actual value of the effect depends on the assumptions made when
the phonon-related features are subtracted.
The relative suppression of the low frequency SW (LSW) evaluated below $\omega _C =$ 800 \wn
becomes more evident when compared with the total SW
(TSW) evaluated up to frequencies of $\omega _C =$ 9500 \wn.  
The inset of Fig. 4 shows the relative changes of the SW with decreasing temperature from 300 K,
expressed in terms of the effective electron density per unit cell
\begin{equation}
N_{eff} = \frac{2 m_e V_{cell}}{\pi e^2}\int\limits^{\omega _C} \sigma (\omega )d\omega,
\end{equation}
where $\sigma (\omega )$ is the IR conductivity after subtraction of
the phonons and the volume of the unit cell ($P4/mmm$) is 
$V_{cell} \simeq  1.7\times 10^{-22}cm^{3}$.
The effective carrier densities with $\omega _C$ = 9500 \wn 
and $\omega _C$ = 800 \wn evaluated at 300 K are 0.12 and 0.005, respectively.
As shown in the inset of Fig. 4, the temperature dependence of the LSW
fully reproduces that of TSW for $300 K\geq T\geq 100 K$. On further decreasing
temperatures the TSW continues to increase, whereas the
LSW remains constant or even slightly decreases.   
We suggest that the discrepancy in the temperature dependences of the LSW and TSW 
reflects the opening of the pseudogap in \ru superconductor. However, the
effect is not very pronounced because of a significant contribution from 
the polaron-like mid-IR band to the LSW. 

The mid-IR band reveals the shape of a small polaron with a clearly resolved
maximum at 4800 \wn at 300 K corresponding to a polaron binding energy of about 0.3 eV.
The most remarkable features of this band are the anomalies which occur
at the onset of ferromagnetic order within the Ru-O layers at
$T_M$ = 133 K, namely, the shift to lower energies from 4800 \wn at $T =180$ K  
to 4400 \wn at $T = 50$ K and a strong increase in the SW.
The inset of Fig. 4 shows that the TSW contributed mainly from the mid-IR band 
increases continuously on decreasing temperature and reveals an anomaly 
at $T_M$, as indicated by a significant deviation from the guideline
which roughly approximates the high-temperature dependence.
The ferromagnetism in Ru-O layers is believed to have the same origin as in SrRuO$_3$,
which, in turn, is explained by itinerant-electron ferromagnetism based
on band theory.\cite{SRO} The Ru ion in SrRuO$_3$ is in a tetravalent 
low-spin state $4d^{4+}: t^3_{2g}\uparrow t^1_{2g}\downarrow e^0_g$ ($S$ = 1).
The ferromagnetic coupling between Ru ions is transmitted via the itinerant
electrons of the minority $t_{2g}$ band, the charge carrier transport being
strongly dependent on the core spin alignment. In \ru the $t_{2g}$ minority band
is partially occupied due to the hole doping of the CuO$_2$ biplanes resulting
in weak itinerant ferromagnetism. Assuming that the mid-IR band is associated with the 
hopping conductivity of the minority $t_{2g}$ electrons, the strong carrier-spin
interaction can naturally explain the temperature behavior of the band.
As T is decreased below $T_M$ the resulting onset of ferromagnetic correlations
leads to an increase of the kinetic energy of the charge carriers, which, 
in turn, results in a decrease of the effective electron-phonon coupling.
As a result, an increase in the TSW and a shift of the polaronic band to lower frequencies
are observed.
Note that this behavior of the SW is typical of magnetoresistive FM compounds like the
doped manganites,\cite{CMR} and has been discussed in terms of a temperature-dependent
competing effects between electron-phonon coupling and double-exchange interaction of
core spins mediated by itinerant electrons.\cite{Millis} 

A further important feature associated with the magnetic transition in \ru
is the anomalous shift of the Ru-related phonon mode at 190 \wn as shown
in Fig. 5(a). In Fig.5(b) we show the relative shifts of the low-frequency phonon modes at 
128, 151, and 190 \wn along with the shift of the apical oxygen phonon
mode at 654 \wn and the broad mid-IR band at 4800 \wn observed on decreasing temperature from 300 K.
Among all phonon modes, only the apical oxygen mode exhibits the classical inharmonic
increase of the eigenfrequency on decreasing temperature, revealing anomalies neither at $T_{SC}$
nor at $T_M$. 
At the same time the low-frequency phonon modes show a characteristic softening at $T_{SC}$
as observed in Y-123 related compounds, and as discussed in terms
of the resonance-frequency renormalization of the phonon modes lying below the
SC gap.\cite{phonons1} However, the most pronounced effect is the anomalous softening
of the mode at 190 \wn which is related to the 
magnetic transition temperature in \ru compound, as shown in Fig. 5.  
The anomalous shift of the phonon mode at 190 \wn, reaching a relative value of 10\% 
at the lowest temperatures, cannot be explained by structural changes
of the bond lengths observed in \ru,\cite{structure2} all being less than
0.5\%. This fact indicates that the Ru-related phonon mode is strongly affected
by the interaction with the electronic system. It is important to note that the shifts
of the phonon mode at 190 \wn and the polaronic band at 4800 \wn exhibit apparently correlated
character, indicating a strong coupling of this phonon mode to the low-energy electronic
excitations, as clearly demonstrated in Fig. 5(b)  
The onset of this effect at the magnetic transition temperature 
and the decrease of the electron excitation energy on lowering temperature
below T$_M$ can be explained by the renormalization of the electron-phonon coupling:     
The kinetic energy of the itinerant electrons increases due to the
onset of the ferromagnetic correlations in the Ru-O layers.   
This scenario can in turn explain the underdoped character of the stoichiometric
\ru superconductor, where the strong hole depletion of the cuprate bilayers is caused by the charge
carrier self-trapping on the Ru moments. 
We should note finally that as seen in Fig 5(b), the 4800 \wn band has the anomaly at $T_M$ on decreasing
temperature with saturation below $T_{SC}$, while the 190 \wn mode exhibits
a singularity at about 90 K and further steep softening associated with the
SC transition. This suggests that the bilayer plasmon-phonon coupling 
responsible for the anomaly of the Cu-O bending mode   
could also play a role in the softening of the 190 \wn mode.
To shed light on the origin of the predominant contribution, the detailed
analysis of the mode eigenvectors is necessary, as well as of the temperature
dependence of the oscillator strength and the damping factor. The accurate
analysis of the temperature dependences will be possible after \ru
single crystals have been prepared. On the other hand, the \ru system could provide an intriguing
opportunity to develop the model of the bilayer Josephson plasmon for the case when the ferromagnetic
moment is in the insulating RuO$_2$ plane mediated the cuprate bilayers. Further investigations
of the phonon anomalies in \ru is promising to study the effect of the Ru magnetic moment
ordering tuned by applying an external magnetic field on the interbilayer coupling. 
\section{Conclusions}
In conclusion, we have investigated the optical conductivity of \ru.
The transformation of the Cu-O bending phonon mode at 288 \wn in the normal
state to an absorption "bump" at 308 \wn in the SC state has been observed.
This observation strikingly resembles data reported recently
for the strongly underdoped YBa$_2$Cu$_3$O$_{6.45}$ crystals, which were
explained with a model invoking the onset of interbilayer and intrabilayer Josephson junctions. 
The present observation evidences a strongly underdoped character 
of the bulk superconductivity originated in weakly coupled cuprate biplanes.
The optical conductivity of \ru is discussed in terms of a two-fluid
model, the far-IR and $dc$ conductivity being predominantly associated with the conductivity
in the CuO$_2$ planes, whereas the mid-IR conductivity being associated with the hopping conductivity
in the Ru-O layers. The correlated anomalous shifts of the Ru-related phonon mode at 190 \wn and the polaron
band at 4800 \wn to lower frequencies have been observed on decreasing temperature below the $T_M$ = 133 K. 
The underdoped character of the stoichiometric \ru superconductor
is explained by the strong hole depletion of the cuprate bilayers caused by the charge
carrier self-trapping on the Ru moments. 
\acknowledgements{This work has been supported by the BMBF/VDI via contract
number EKM/13N6917, partly by the Deutsche Forschungsgemeinschaft via the
Sonderforschungsbereich 484 (Augsburg) and partly by the DAAD.}

\newpage
\begin{figure} [t]
\centerline{\epsfxsize=3.375in\epsfbox{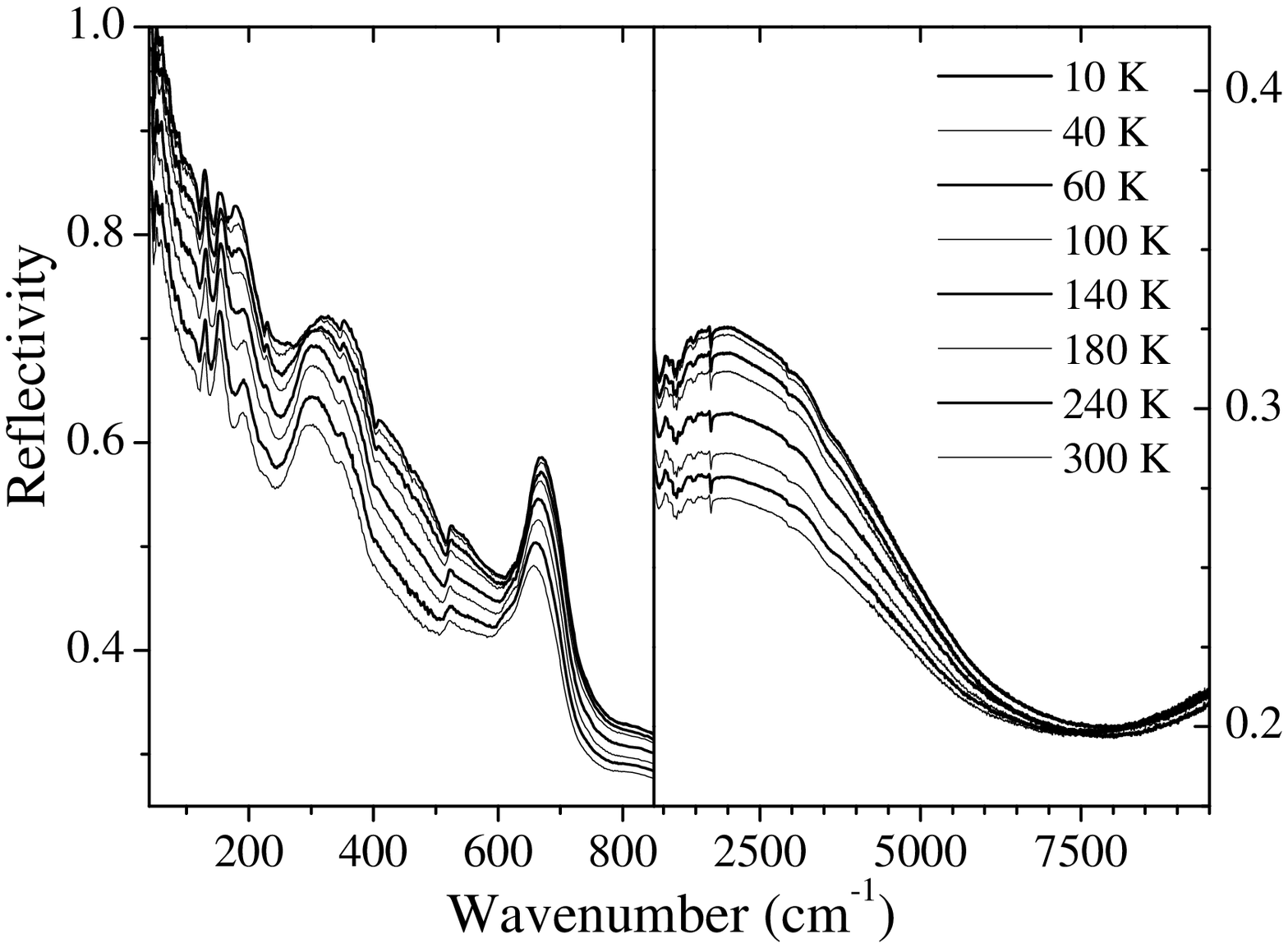}}
\caption{Reflectivity spectra of \ru at different temperatures. 
Note the change of scales from the left to the right panel.}
\label{Fig1}
\end{figure}
\newpage
\begin{figure} [t]
\centerline{\epsfxsize=3.375in\epsfbox{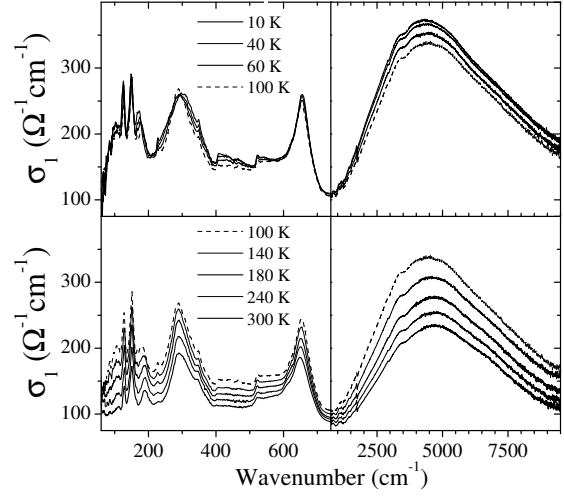}}
\caption{Real part of the optical conductivity of \ru with $T_{SC}$= 45K
and $T_{FM}$= 133K. Spectra are shown for T = 10, 40, 60, and 100 K - upper
panels, and 100, 140, 180, 240, and 300 K - lower panel. Note the change
of wavenumber scale from left to right panels.}
\label{Fig2}
\end{figure}
\newpage
\begin{figure} [t]
\centerline{\epsfxsize=3.375in\epsfbox{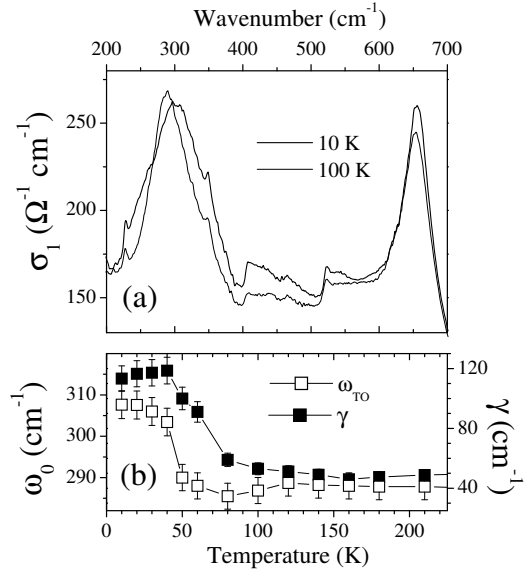}}
\caption{(a) Real part of the optical conductivity at high-phonon frequencies
at two different temperatures well above and well below $T_{SC}$= 45 K.
(b) Temperature dependence of the position and the Lorentzian damping of the asymmetric
peak at 288 \wn (see text).}
\label{Fig3}
\end{figure}
\newpage
\begin{figure} [t]
\centerline{\epsfxsize=3.375in\epsfbox{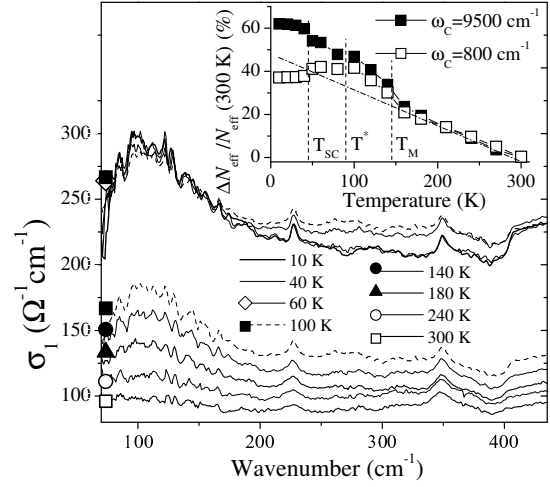}}
\caption{Temperature  dependence of the electronic background far-IR conductivity
with phonons being subtracted by fits to oscillators as described
in the text. The curves for T = 10, 40, 60, and 100 K are shifted up
by 100 $\Omega ^{-1} cm ^{-1}$ for clarity. The symbols on the Y-axis represent the
$dc$ conductivity values. Inset: Temperature dependences of $N_{eff}$ evaluated
below $\omega _C$ and related to those at T = 300 K. The dot-dashed guideline continues
the temperature dependences from above to below T$_M$.}
\label{Fig4}
\end{figure}
\newpage
\begin{figure} [t]
\centerline{\epsfxsize=3.375in\epsfbox{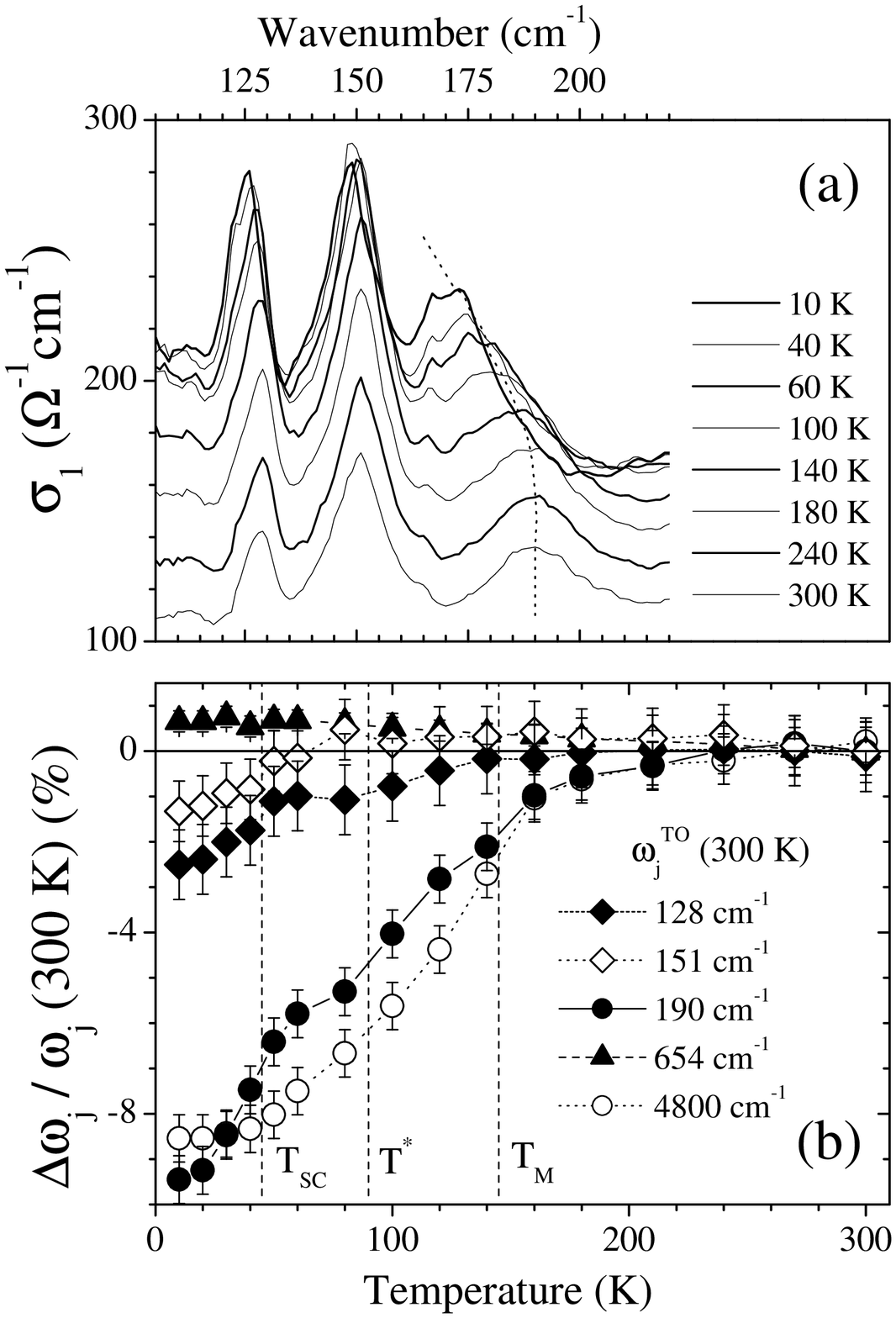}}
\caption{(a) Real part of the optical conductivity at low phonon frequencies
at different temperatures. The dotted line shows the temperature dependence
of the eigenfrequency of the Ru-related mode derived by the dispersion
analysis. (b) The relative shift of the resonant 
frequencies of the phonon modes and the mid-IR band observed on decreasing temperature from 300 K.
The vertical dashed lines mark the critical temperatures. }
\label{Fig5}
\end{figure}

\begin{references}
\bibitem{Bmuon}  C. Bernhard {\it et al.},
Phys. Rev. B{\bf 59}, 14099 (1999).
\bibitem{Bmeissner}C. Bernhard, J. L. Tallon, E. Br\"{u}cher and R. K. Kremer,
Phys. Rev. B{\bf 61}, R14960 (2000).
\bibitem{structure1}A. C. McLaughlin {\it et al.},
Phys. Rev. B{\bf 60}, 7512 (1999).
\bibitem{structure2}  O. Chmaissam {\it et al.},
Phys. Rev. B{\bf 61}, 6401 (2000).
\bibitem{Btransport}J. L. Tallon, J.W. Loram, G. V. M. Williams, C. Bernhard,
Phys. Rev. B{\bf 61}, R6471 (2000).
\bibitem{AFMorder}J. W. Lynn {\it et al.},
Phys. Rev. B{\bf 61}, R14964 (2000).
\bibitem{FMorder}G. V. M. Williams, S. Kr\"{a}mer, Phys. Rev. B{\bf 62}, 4132 (2000).
\bibitem{plasmon1}M. Gr\"{u}ninger, D. van der Marel, A. A. Tsvetkov, and
A. Erb, Phys. Rev. Lett. {\bf 84}, 1575 (2000).
\bibitem{noplasmon}L. B. Ioffe and A. J. Millis, Phys. Rev. B {\bf 61}, 9077 (2000).
\bibitem{plasmon2}D. Munzar {\it et al.},
Solid State Commun. {\bf 112}, 365 (1999).
\bibitem{anomaly3}C. Bernhard {\it et al.},
Phys. Rev. B {\bf 61}, 618 (2000).
\bibitem{sample}J. Hemberger {\it et al.}, 
preprint.
\bibitem{ir}A. P. Litvinchuk {\it et al.},
Phys. Rev. B {\bf 62}, 9709 (2000).
\bibitem{phonons1}A. P. Litvinchuk, C. Thomsen, and M. Cardona, in {\it
Physical Properties of High Temperature Superconductors IV}, edited by
D. M. Ginsberg (World Scientfic, Singapore, 1994), p. 375 and references
therein.
\bibitem{phonons2}R. Henn, T. Strach, E. Sch\"{o}nherr, and M. Cardona,
Phys. Rev. B {\bf 55}, 3285 (1997).
\bibitem{phonons3}J. Sch\"{u}tzmann {\it et al.},
Phys. Rev. B {\bf 52}, 13665 (1995).
\bibitem{Fano}A. Damascelli, K. Skulte, and D. van Marel, A. A. Menovsky,
Phys. Rev. B {\bf 55}, R4863 (1997).
\bibitem{anomaly1}C. C. Homes {\it et al.},
Physica C {\bf 254}, 265 (1995).
\bibitem{anomaly2}R. Hauff, S. Tajima, W.-J. Jang, and A. I. Rykov, Phys. Rev. Lett. {\bf 77}, 4620 (1996).
\bibitem{pseudogap}Tom Timusk and Bryan Statt, Rep. on Progress in Phys. {\bf 62}, 61 (1999).
\bibitem{Zhu}Jian-Xin Zhu, C.S. Ting, and C.W. Chu, Phys. Rev. B {\bf 62}, 11369 (2000).
\bibitem{SRO} S.C. Gausepohl {\it et al.}, Phys. Rev. B {\bf 52}, 3459 (1995);
I.I. Mazin, D.J. Singh, Phys. Rev. B {\bf 56}, 2556 (1997); J.S. Dodge
{\it et al.}, Phys. Rev. B {\bf 60}, R6987 (1999).
\bibitem{CMR}A.V. Boris {\it et al.},
Phys. Rev. B {\bf 59}, R697 (1999).
\bibitem{Millis} A.J. Millis, R. Mueller, and Boris. I. Shraiman, Phys. Rev. B {\bf 54}, 5405 (1996);
A. Chattopadhyay, A.J. Millis, S. Das Sarma, Phys. Rev. B {\bf 61}, 10738 (2000).
\end{references}
\end{document}